# Exploiting the optical quadratic nonlinearity of zinc-blende semiconductors for guided-wave terahertz generation: a material comparison

Matteo Cherchi[1,2*], Alberto Taormina[3†], Alessandro C. Busacca[1,3], Roberto L. Oliveri[1,2], Saverio Bivona[1,2], Alfonso C. Cino[3,4], Salvatore Stivala[3], Stefano Riva Sanseverino[3,4], and Claudio Leone[1,2]

[1]CNISM – Consorzio Nazionale Interuniversitario per le Scienze Fisiche della Materia, Unità di ricerca di Palermo, Università di Palermo, Piazza Marina 57, I-90133 Palermo
[2]DIFTER – Dipartimento di Fisica e Tecnologie Relative, Università di Palermo, viale delle Scienze edificio 18, I-90128 Palermo, Italy
[3]DIEET- Dipartimento di Ingegneria Elettrica, Elettronica e delle Telecomunicazioni, Università di Palermo, viale delle Scienze edificio 9, I-90128 Palermo, Italy
[4]CRES - Centro per la Ricerca Elettronica in Sicilia, Via Regione Siciliana 49, I-90046 Monreale (PA), Italy
[*]Corresponding author, cherchi@unipa.it
[†]Current address: Laboratoire Matériaux et Phénomènes Quantiques, CNRS–UMR 7162, Université Paris–Diderot, bâtiment Condorcet, Bureau 677A 10, rue Alice Domon et Léonie Duquet, 75205 Paris cedex 13, France

*Abstract* - **We present a detailed analysis and comparison of dielectric waveguides made of CdTe, GaP, GaAs and InP for modal phase matched optical difference frequency generation (DFG) in the terahertz domain. From the form of the DFG equations, we derived the definition of a very general figure of merit (FOM). In turn, this FOM enabled us to compare different configurations, by taking into account linear and nonlinear susceptibility dispersion, terahertz absorption, and a rigorous evaluation of the waveguide modes properties. The most efficient waveguides found with this procedure are predicted to approach the quantum efficiency limit with input optical power in the order of kWs.**

## Introduction

In recent years there has been a lot of effort to provide more practical terahertz sources for out-of-laboratory applications [1]-[3], because of the large number of potential applications for security, medical imaging, pharmaceutical industry, semiconductor industry, etc.[4]. Terahertz science is a meeting point of two completely different and somewhat complementary sciences: the science of ultra high frequency microwave circuits and the science of far infrared optics.

Many different approaches have been proposed and implemented for terahertz generation, including free electron lasers [5], gas lasers [6], silicon lasers [7], quantum cascade lasers [8], linear and nonlinear microwave circuits [9]. Approaching the terahertz range from the optical side can be advantageous due to the availability of low cost laser sources and of nonlinear materials suitable for frequency down-conversion. In fact, a very promising approach for terahertz generation relies on the difference frequency generation (DFG) process - often reported also as optical rectification when using a single broadband optical source [10]- in quadratic nonlinear materials [11]. In DFG experiments, these materials are lit up with photons of angular frequencies $\omega_u$ and $\omega_v$, that are the so called pump photons and signal photons, such that $\omega_u > \omega_v$. The nonlinear matter-radiation interaction can turn a pump photon into a signal photon, with the further effect of creating an additional photon with frequency $\omega_w = \omega_u - \omega_v$, as required by energy conservation. Clearly, the higher the optical intensity, the higher the probability of this process to occur.

The wide availability of tunable laser sources and the relatively simple optical configurations required, counterbalance the small overall energetic efficiency which can be achieved with DFG even at the quantum limit, due to the large photon energy separation. Generation efficiency is additionally limited by large phonon absorption in the terahertz domain and linear or nonlinear absorption in the optical domain, by the amount of quadratic nonlinearity and by the optical damage threshold of the considered material. Furthermore, in the case of broadband optical pumping, conversion efficiency may be affected by destructive interference among DFG processes at different wavelengths [12]. Anyway, the main limitation comes from phase mismatch between the two optical waves and the terahertz wave: due to the different phase velocities of the three waves, in general terahertz photons created at different propagation steps interfere destructively. Therefore, for efficient terahertz generation, phase matching is mandatory. As a consequence of all of these limitations, most results reported in the literature lie far below the quantum efficiency limit. Only in a recent paper [13], the quantum efficiency limit was eventually approached using DFG (up to 39.2% photon conversion), but this was accomplished at the expense of launching megawatts peak power pulses on bulk GaSe, a birefringent nonlinear material with very low terahertz absorption.

As a matter of principle, for this kind of applications relevant advantages should be expected from many zinc-blende compounds, like GaAs, InP, GaP and CdTe, that are well known for their very good quadratic nonlinear optical properties [14]. Nevertheless, apart from electro-optic applications, these nonlinear properties cannot be easily exploited, because the intrinsic isotropy of zinc-blende refractive index does not allow birefringent phase matching [15]. To overcome this limit, quasi phase matching [15], [16], plasmon phase matching [17], surface grating and prism phase matching [18], small-angle non-collinear phase matching [19], and modal phase matching [20], [21] have been proposed and experimentally demonstrated. Nevertheless, for frequency mixing applications in the optical domain, these alternative solutions can hardly compete with the ease of birefringent phase matching in anisotropic crystals. On the contrary, in the case of optical DFG in the terahertz domain, modal (waveguide) phase matching can become an advantageous alternative to birefringent phase matching. This is because some of the aforementioned zinc-blende crystals feature very large nonlinear coefficients and acceptable absorption coefficients in the terahertz range, but also because, in the guided-wave configuration, high light intensities can be achieved with orders of magnitude smaller laser pump power and over much longer interaction lengths than in bulk configurations.

**Modal phase matching**

In a pioneering work [22] published in 1974, guided-wave phase matched terahertz generation from $CO_2$ laser lines was shown in planar GaAs waveguides. More recently some papers have theoretically studied the case of three dimensional (3D) confined modes [17], [23]-[30] and some of these also reported experimental results [28]-[30]. Anyway all of these papers neither addressed the problem of conversion efficiency optimization nor the problem of possibly reaching the quantum efficiency limit.

In this paper we propose a general method to determine the best 3D guided-wave configurations by taking into account both material and modal properties. We chose the simpler rectangular channel waveguide configuration with air cladding (however the method applies to any waveguide geometry) and, thanks to a suitable Figure of Merit (FOM) definition, we found the best performing waveguides for terahertz generation by DFG. For the optical pump and signal wavelengths, we chose the near infrared region around 1.55 μm wavelength, not only because this is covered by many different laser sources, but also because it falls in the transparency window of the most common zinc-blende semiconductors. This choice clearly means that the waveguide can be nearly single mode for the terahertz radiation only, while, in practice, it will act as a bulk material for the optical radiation with tens to hundreds times smaller wavelengths. At variance to other works [25], [26], [29], we didn't strive to ensure single mode condition also for the optical radiation, in which case the light intensity in the optical mode would be orders of magnitude greater than the intensity available to the DFG process. In turn, this would mean wasting the maximum launchable optical peak power allowed by the material damage threshold. Our choice is also supported by numerical simulations showing that, with a suitably shaped collimated Gaussian beam, optical coupling to the input section of our highly multimodal waveguides can almost equal the plane waves Fresnel limit. This is done with high tolerance to misalignments, and with high suppression of coupling to higher order modes.

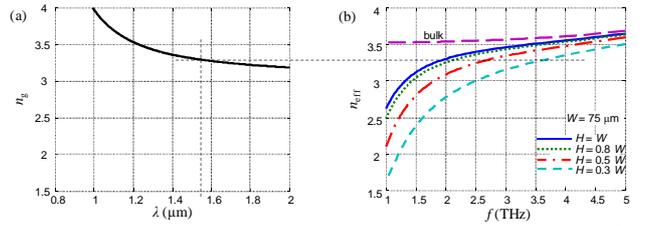

**Fig. 1. Graphical determination of the phase matching condition.** On the left: wavelength dependence of the optical group index $n_g$ of InP. On the right: dispersion curves of rectangular InP waveguides surrounded by air with a fixed value of the waveguide width $W$ and different waveguide heights $H$. The chosen optical wavelength is 1.55 μm. Phase matching is ensured whenever the optical group index $n_g$ equals the terahertz effective index $n_{eff}$. Notice that phase matching could never be attained using this pump wavelength in the bulk material.

Within the waveguide, we know also that, for pump frequencies much greater than the generated frequency, the phase matching condition is accurately approximated by the equality between optical group index $n_g$ and terahertz effective index $n_{eff}$ [24], [31]. In order to determine the relationship between the phase matching frequency and the waveguides cross-sections, we calculated the optical group index, and both terahertz refractive index and absorption coefficient from the experimental data reported in the literature [32]. Then, in order to calculate effective indexes, modal losses, and spatial distributions $e_q(x, y)$ $(q = w, v, u)$ of the waveguides modes, wewe resorted to a fully vectorial commercial modal solver [33] based on field mode matching [34], that is a very efficient and reliable semi-analytical method for rectangular step index waveguides, even in the case of high index contrast. Since the refractive index of zinc-blende crystals in the terahertz region is typically greater than 3, turning to this mode solver overcomes a critical point of some previous papers [17], [23], [28] making use of scalar approximated methods, that are not reliable [35] when dealing with such a high index contrast single mode waveguides, due to the effects at the waveguide edges. Comparing these approximated methods with our rigorous numerical results, we found waveguide size errors even greater than 50%.

As a practical procedure, we fixed a waveguide width $W$ and then calculated the dispersion curves of the terahertz fundamental TE modes (i.e. those having the main electric field component along the $W$ direction) for different waveguide height $H$, as shown in Fig. 1.b for the case of InP. In this way, it is straightforward to find the phase matched frequencies in the terahertz domain, by comparison with the group index in the optical domain shown in Fig. 1.a. It is clear that the phase matching condition can be found only when the chosen optical pump wavelength corresponds to an optical group index smaller than the terahertz bulk refractive index. It is also clear from Fig. 1 that, for any given terahertz frequency $f_0$, there always exists one and only one optical wavelength $\lambda_0$ ensuring phase matching in the bulk material. On the contrary, the advantage of choosing a guided-wave configuration is that we can be freed, to a certain extent, from the material properties constraints. Introducing the waveguide geometry as an additional degree of freedom, the terahertz effective index can be lowered at will, so that DFG at $f_0$ can be phase matched for any fixed optical wavelength $\lambda > \lambda_0$ of practical interest. Furthermore, the greater the

$$A_q \equiv \frac{\langle e_q | e_q \rangle^2}{\langle e_q^2 | e_q^2 \rangle}. \qquad (2)$$

phase mismatch, the greater the required terahertz effective index change, i.e. the smaller the waveguide size. This means that, for a given optical pump power, light intensity will be enhanced together with nonlinear effects. So, in a sense, we could say that guided-wave phase matching turns a problem into an advantage.

## Material properties

In order to move away as possible the damage threshold, that is to allow for the highest possible light intensities, absorption in the optical domain must be kept at a minimum. This is why we cannot consider here materials like InAs, InSb, GaSb, despite their high quadratic nonlinearity: at the optical wavelengths around 1.55 μm we chose to work with, they suffer from excessive absorption. The laser-induced surface-damage threshold in the transparency window of the considered zinc-blendes typically exceeds tens of TW/m² for nanosecond pulses [36], depending on the quality of the crystal growth, and it scales roughly with the inverse of the pulse duration, i.e. with constant pulse energy. We have already noticed that, for a given launched peak power, the smaller the waveguide cross-section, the higher the optical intensity. This is a clear advantage from an efficiency point of view, but it can be seen also as a disadvantage from an energetic point of view, since the intensity damage threshold clearly limits the highest launchable peak power. Another important point is that, in quadratic mixing of three waves propagating in their own waveguide modes, the effective light intensity for the nonlinear process must be calculated over the effective area [37]

$$A_{DFG} \equiv \frac{\langle e_w | e_w \rangle \langle e_v | e_v \rangle \langle e_u | e_u \rangle}{\langle e_w e_v | e_u \rangle^2}, \qquad (1)$$

that is the inverse of the DFG overlap integral, accounting for the overlap of the spatial distributions $e_q(x,y)$ of the main electric field component of each mode. Here we resorted to the usual definition of the scalar product of spatial distributions over the waveguide cross-section spatial coordinates $x$ and $y$

$$\langle f | g \rangle \equiv \int_{-\infty}^{+\infty} \int_{-\infty}^{+\infty} f^*(x,y) g(x,y) dx\, dy.$$

| Material | $f_{TO}$ (THz) | $d_{36}$ (pm/V) | $C$ |
|---|---|---|---|
| GaAs | 8.1 | 83÷209 | -0.55 |
| InP | 9.1 | 143÷287 | -0.52 |
| GaP | 10.9 | 53÷218 | -0.51 |
| CdTe | 4.2 | 59÷170 | -0.5 |

**Table 1. Material properties.** Frequencies of the TO phonon resonances and experimental data ranges of the electronic contribution to the quadratic nonlinear coefficient $d_{36}$ for the materials under study.

On the other hand the usual linear light intensity for each mode must be calculated over the effective areas:
In the case of waveguides designed to be single mode also in the optical domain, the poor match between terahertz and optical modal shapes makes $A_{DFG}$ always comparable with the effective area of the terahertz mode. As a result, the effective light intensity available to the nonlinear process comes out to be much smaller than the light intensity launched in the small optical mode.

As regards terahertz absorption, this is never negligible at room temperature in any crystal, mainly due to the transverse optical phonon resonance, whose frequency $f_{TO}$ typically fall in the (4÷11) THz range [32] for zinc-blende crystals (see Table 1). This means that the material contribution to terahertz conversion efficiency will be mainly governed by the terahertz absorption coefficient $\alpha$ and the quadratic nonlinear susceptibility $\chi^{(2)}_{ijk}$. Despite the tensorial nature of the latter, the symmetry of zinc-blende crystals reduces the number of its independent quadratic nonlinear elements to one [11], so that $\chi^{(2)}_{ijk} = 2 d_{36} \times |\varepsilon_{ijk}|$. In practice, after a proper choice of the crystal orientation, we can choose the more suitable polarizations of the electric fields $E_u$, $E_v$, and $E_w$, corresponding to the launched optical pump and optical signal and to the generated terahertz wave respectively. In this way it is straightforward to associate an effective scalar nonlinear coefficient $d_{eff}$ to any chosen configuration [15]. For example, in Fig. 2 two convenient choices are presented, the first one corresponding to $d_{eff} = 2/\sqrt{3} d_{36}$ and the second one corresponding to $d_{eff} = d_{36}$. The calculations in this paper assume the first configuration.

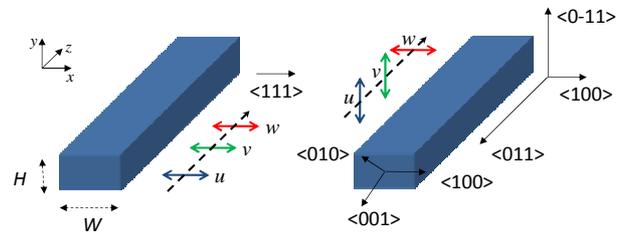

**Fig. 2. Two possible configurations.** Two of the possible choices for the orientations of the crystal axes and for the linear polarizations of the optical pump $u$, optical signal $v$ and terahertz wave $w$. The configuration on the left is the most efficient.

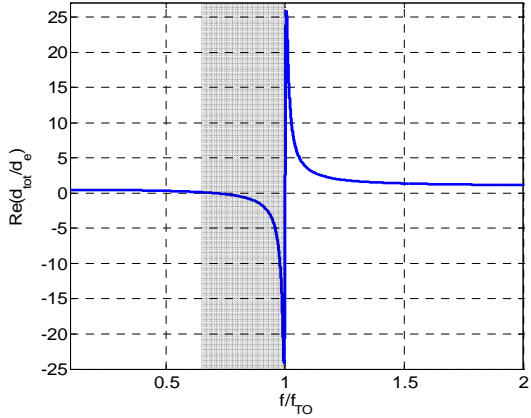

**Fig. 3. Oscillator model for the nonlinear coefficient**. Normalized dispersion of the nonlinear coefficient vs. normalized generated frequency for terahertz generation by optical DFG, assuming a Faust-Henry coefficient $C = -0.5$ and a resonance linewidth $\gamma = 10^{-2} f_{TO}$.

From a comparison of the nonlinear coefficients data available in the literature [14], [15], [38], [39],, we can restrict our analysis to the most promising III-V crystals GaAs, InP, and GaP, and to the II-VI crystal CdTe (see Table 1). At a first glance it is clear that the conversion efficiency in crystals like CdTe is much more affected by terahertz absorption because the phonon resonance lies in the very middle of the terahertz region. Also notice that the nonlinear coefficients reported in the literature are usually referred to the optical domain, thus they should be used with caution in situations like those considered here. In an early paper [40], Faust and Henry experimentally showed that, when quadratic frequency mixing involves one frequency near the phonon resonance $f_{TO}$, and two frequencies well above it, the nonlinear coefficient $d_{tot}$ is determined by the sum of the standard electronic contribution $d_e$ measured in the optical domain, and a lattice contribution resonating at $f_{TO}$. They introduced the so called Faust-Henry coefficient $C$, measuring the ratio between these two contributions, in order to write the resulting nonlinear coefficient in the form

$$d_{tot}(f) = \text{Re}\left\{d_e \left[1 + \frac{C}{D(f)}\right]\right\} \quad (3)$$

where $D(f) \equiv 1 - (f/f_{TO})^2 - if\gamma/f_{TO}^2$, $f$ being the lowest frequency involved in the process. In experiments on zinc-blende crystals at room temperature, the coefficient $C$ turns out to be a negative number ranging from $-0.6$ to $-0.2$ [41]. As shown in Fig. 3, since typically $\gamma \approx 10^{-2} \times f_{TO}$, the zeros of Eq. (3) correspond approximately to $f_{TO}$ and $\tilde{f} \equiv \sqrt{1+C} \times f_{TO}$. Between these two frequencies (shaded region) $d_{tot}$ changes its sign reaching its minimum value $d_e(1 + C/2 \times f_{TO}/\gamma)$, at frequency $f_{TO} - \gamma/2$, while its maximum $d_e(1 - C/2 \times f_{TO}/\gamma)$ corresponds to frequency $f_{TO} + \gamma/2$. Notice also that, for frequencies $f < \tilde{f}$, the nonlinear coefficient is always reduced by a factor smaller than $(1 + C)$. For the III-V materials under study, we used the $C$ values reported in Ref. [41]. As shown in Ref. [42], the $C$ value for CdTe can be derived from data available in the literature [43] to be around -0.5. We have also assumed $\gamma = 10^{-2} \times f_{TO}$ for all materials.

We stress, at the same time, the wide spread of nonlinear coefficient data for the materials under study and the lack of detailed (i.e. beyond the simple oscillator model fits found in Ref. [32]) experimental terahertz absorption data for some of them. In fact all of these uncertainties can greatly affect our theoretical calculations. As a rule we assumed the highest reported values of the nonlinear coefficients to be the reference ones and the absorption data from the oscillator models to be reliable, unless otherwise specified. Anyway, whenever these assumptions should reveal to be wrong, it will be straightforward to rescale the found results accordingly.

## Definition of a figure of merit

At a given waveguide section $z$, we can write the main electric field component of the three interacting guided wave modes with angular frequencies $\omega_u > \omega_v \gg \omega_w$ (so that $\omega_w = \omega_u - \omega_v$) and propagation constants $k_q$ ($q = u, v, w$) as [44],[45]

$$E_q(x, y, z; t) = \frac{1}{2}\mathcal{E}_q(x, y, z; t) \exp[i(\omega_q t - k_q z)] + \text{c.c.} \quad (4)$$

where

$$\mathcal{E}_q(x, y, z; t) \equiv \sqrt{\frac{2Z_0 \hbar \omega_q}{n_q \langle e_q | e_q \rangle}} e_q(x, y) q(z; t). \quad (5)$$

Here $\hbar$ is the Planck constant, $Z_0$ is the vacuum impedance and $n_q$ are the effective indexes of the three waveguide modes with spatial distributions $e_q(x, y)$, while the $q(z; t)$ functions are assumed to be slowly varying functions of $z$ and are normalized such that their square moduli are the photon fluxes (number of photons per unit time) in each mode.

Under these assumptions, the slowly varying envelope approximation [44] applies, thus quadratic three-wave mixing with terahertz absorption can be modeled by the following coupled differential equations for the $q(z;t)$ functions

$$\begin{cases} \dfrac{dw}{dz} = -i\xi uv^* e^{-i\Delta kz} - \dfrac{\alpha}{2}w \\ \dfrac{dv}{dz} = -i\xi uw^* e^{-i\Delta kz} \\ \dfrac{du}{dz} = -i\xi vw e^{+i\Delta kz} \end{cases} \quad (1)$$

where the momentum mismatch $\Delta k \equiv k_u - k_v - k_w$ is in general non-zero and we have defined

$$\xi \equiv d_{\text{eff}} \sqrt{\dfrac{2Z_0 \hbar \omega_w \omega_v \omega_u}{c^2 n_w n_v n_u A_{\text{DFG}}}}, \quad (2)$$

where $c$ is the vacuum speed of light. Notice that we are also assuming terahertz absorption lengths much longer than terahertz wavelengths [46], negligible optical losses, and pulse durations not too smaller than the time of flight in the waveguide, in order to avoid the effects of group velocity dispersion.

A closed-form solution for Eqs. (1) is not available but in the unrealistic cases of negligible terahertz losses or equal losses in all of the three modes [16]. Anyway, since we are interested in phase matched processes, we can set $\Delta k = 0$, and then notice that the resulting equations are form invariant under the rescaling transformation:

$$\begin{cases} \xi \to r \times \xi \\ \alpha \to r \times \alpha \\ z \to r^{-1} \times z \end{cases}, \quad (3)$$

where $r$ is any real number. This suggests to define a novel FOM

$$\mathcal{F} \equiv \dfrac{\xi^2}{\alpha^2} = \dfrac{2\hbar \omega_w \omega_v \omega_u}{c^3 \varepsilon_0 A_{\text{DFG}}} \times \dfrac{d_{\text{eff}}^2}{n_w n_v n_u \alpha^2} \equiv \quad (4)$$
$$\mathcal{F}_W \times \mathcal{F}_M$$

as a measure of the photon conversion capability of the whole system. In Eq. (4) we have highlighted two factors. The first one, $\mathcal{F}_W$, is a waveguide FOM, depending on the optical wavelengths, on the terahertz frequency, and on the chosen waveguide geometry ensuring phase-matching. The second one, $\mathcal{F}_M$, (already proposed by Ding [13]) is a material FOM, that accounts for the linear and nonlinear properties of the chosen material: the phase matching condition requires $n_w$ to equal the optical group index, while the highly multimodal nature of the waveguide in the optical domain makes, in practice, both $n_v$ and $n_u$ equal the bulk refractive index of the waveguide core.

The aforementioned mathematical invariance physically means that any two structures - which in general can be made of different materials, can have different cross-sections, can be pumped with different optical wavelengths, and can be meant to generate different terahertz frequencies - must behave in the same way whenever they feature the same $\mathcal{F}$ value, up to a length rescaling proportional to the absorption length. This makes Eq. (4) a meaningful FOM definition. Actually, this is a very general definition, since it applies to any phase-matched three wave mixing system – either guided-wave or free propagating- damped by a single absorption coefficient, no matter how phase matching is achieved. This means that $\mathcal{F}$ makes it possible comparisons between any kind of phase matched terahertz generation systems. It is also worth noticing that $\mathcal{F}$ is proportional to the product of all the three frequencies involved in the process, and so the shorter the optical pump wavelength, the easier the photon conversion. On the other hand, from an energetic point of view, the Manley-Rowe relations impose that energy conversion efficiency, defined as the ratio between generated terahertz power $P_w$ and total launched optical power $P_{u0} + P_{v0}$, must be lower than $\omega_w/\omega_u$, and so longer pump wavelengths are energetically advantageous. This is to point out that the proposed FOM is not a measure of energy conversion capability but only of photon conversion capability.

## Material comparison

Before comparing the FOMs of different waveguides, we now focus on Fig. 4, showing the dependence of $\mathcal{F}_M$ on terahertz frequency for the materials under study, together with the absorption curves used in the calculations. The resonator model accounting for the nonlinear susceptibility dispersion (see Fig. 3) predicts a sign change of the nonlinear susceptibility in the shaded regions, so that the electronic and the ionic contributions cancel each other at the region borders. Hence, in all cases, we can highlight three different frequency regions. The first one corresponds to $f > f_{\text{TO}}$, where modal phase matching is not possible, because the refractive indexes of all these materials are always lower than the optical group index. The second one corresponds to the shaded region $\tilde{f} < f < f_{\text{TO}}$, where terahertz absorption is very high, possibly implying not negligible sample heating, and the refractive index can be even higher than 12. Also nonlinear susceptibilities of all orders are enhanced, even in their imaginary parts, i.e. those responsible for nonlinear absorption. Since modal phase matching with very high refractive indexes requires very small waveguides, the huge index contrast and the very high "squeezing" of the terahertz field make the modal effective index quite sensitive to fabrication errors. Furthermore, in this regime, the study of the nonlinear process cannot be reliably modeled by Eq. (1), because higher order derivatives must be included [46] together with nonlinear cubic terms (self and cross-phase modulation, four wave mixing,…), $\chi^{(2)}$ cascading terms [45], and nonlinear absorption terms [48]. The most promising region corresponds to $f < \tilde{f}$, where terahertz absorption is relatively low. Actually, it should be noticed that, in the literature, measurements of terahertz absorption has been reported for CdTe, GaP and GaAs [32]. The oscillator model for CdTe fits quite well the experimental data even far from resonance, so we used this model in our simulations. In the case of GaP and GaAs the oscillator model works well near the phonon resonance, but underestimates up to two orders of magnitude the experimental data below resonance. So, in this spectral region, we used instead a numerical interpolation of the tabulated data. Unfortunately for InP we couldn't find any experimental terahertz absorption data in the literature, so we were forced to use the (probably optimistic) extrapolated estimations of the oscillator model. For example, the oscillator models for CdTe, GaP, GaAs and InP at 1 THz give absorption coefficients 4.81 cm$^{-1}$, 0.04 cm$^{-1}$, 0.13 cm$^{-1}$ and 0.22 cm$^{-1}$ respectively, while experimental data for CdTe, GaP and GaAs are 4.55 cm$^{-1}$, 1.92 cm$^{-1}$ and 0.48 cm$^{-1}$ respectively. Also it should be noticed that, far from the phonon resonance, terahertz absorption can b significantly affected by other mechanisms, like free-carrier absorption, that, in principle, can be compensated by different dopings. Clearly all of these things this must be taken into account when comparing these materials. Nevertheless, Fig. 4 clearly suggests that III-V materials can perform much better than CdTe and that they can be effectively exploited for DFG below 6 THz.

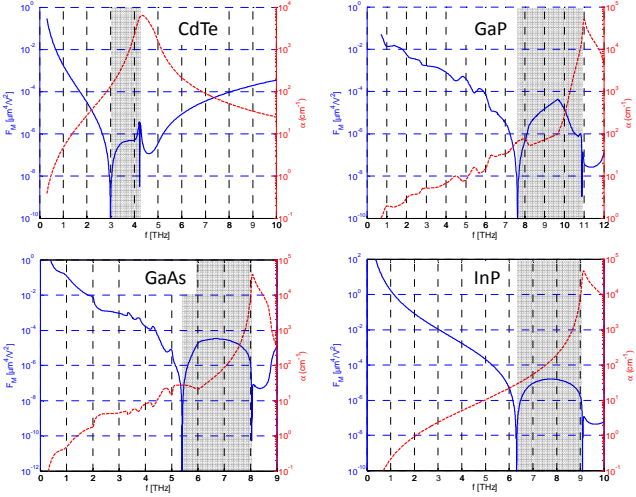

**Fig. 4 Summary of the material properties.** Material figure of merit $\mathcal{F}_M$ (solid lines) and terahertz absorption curves (dashed lines) of the four materials under study. In the shaded regions the second order susceptibility changes its sign.

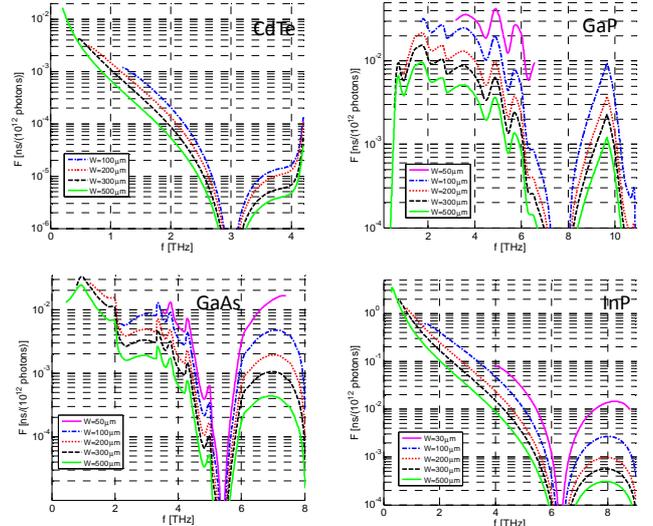

**Fig. 5 Comparison of different waveguides.** Figure of merit $\mathcal{F}$ of phase matched waveguides with different cross sections. Each curve corresponds to a fixed waveguide width $W$. For each curve the higher the phase matching frequency $f$ the smaller is the waveguide height $H$. Best conversion efficiencies are predicted for lower frequencies. At a given frequency, aspect ratios $H/W$ closer to 1 are preferable.

Eventually we notice that the simulated terahertz waveguide losses differ by a few percent only from the bulk material losses, and so they are much smaller than the uncertainties in the experimental absorption data. Therefore we decided to neglect these small corrections in our calculations.

## Comparison of different waveguides

At the light of the previous analysis, we now focus on the whole FOM $\mathcal{F}$, that has the dimensions of the inverse of a photon flux. As shown in Fig. 1, for a fixed waveguide width $W$, it is possible to find a one to one relationship between the phase-matched terahertz frequency $f$ and the waveguide height $H$. In this way we can plot in Fig. 6.a the frequency dependence of $\mathcal{F}$ for different materials and different fixed widths. By a comparison with Fig. 4 it is clear that the waveguide FOM $\mathcal{F}_W$ partially counterbalances the $\mathcal{F}_M$ decrease with frequency. This is because $\mathcal{F}_W \propto f/A_{DFG}$ and $A_{DFG}$ decreases with frequency, spanning two orders of magnitudes in the region of interest. This is a clear additional advantage of choosing the guided-wave configuration. Also it is clear that, for a given terahertz frequency, the choice of the best waveguide cross-section does matter and, as a general rule, aspect ratios $H/W$ closer to 1 are preferable. Once again we stress the need for accurate and reliable experimental data of the absorption coefficients and of the nonlinear coefficients for actual design purposes. Assuming that the experimental absorption data for CdTe, GaP, GaAs are reliable, the best choice among these materials seems to be GaP, but in the region around 1 THz, where GaAs should work better. As regards InP, we notice that it could still have very good performances, even assuming an absorption coefficient two or three times larger than that predicted by the oscillator model. Also, about 1 THz, InP could be the best material even assuming an absorption coefficient one order of magnitude larger.

| Material | $W\times H\times L$ ($\mu m\times\mu m\times cm$) | $\lambda_v$ ($\mu m$) | $f$ (THz) | $\mathcal{F}$ (ns/10$^{12}$) | $\mathcal{F}_M$ ($\mu m^4/V^2$) | $I_u$ (TW/m$^2$) | $I_{DFG}$ (TW/m$^2$) | $P_w$ (W) | $\eta_Q$ (%) |
|---|---|---|---|---|---|---|---|---|---|
| GaP | 100×67.4×0.7 | 1.566 | 2 | 2.5×10$^{-2}$ | 2.8×10$^{-3}$ | 0.48 | 0.50 | 8.7 | 42.0 |
| GaP | 50×18.8×0.2 | 1.591 | 5 | 3.7×10$^{-2}$ | 4.0×10$^{-4}$ | 4.7 | 3.7 | 25.0 | 48.3 |
| GaAs | 300×203.8×3.6 | 1.558 | 1 | 3.2×10$^{-2}$ | 1.2×10$^{-1}$ | 6.2×10$^{-2}$ | 6.5×10$^{-2}$ | 4.8 | 46.6 |
| InP | 200×115×1.5 | 1.558 | 1 | 0.98 | 1.6 | 0.14 | 0.16 | 9.0 | 86.6 |

**Table 1. Results of numerical simulations.** Predicted terahertz generation in some of the optimal waveguides analyzed in Fig. 5. In all cases 2 kW pump power and equal signal power have been assumed. Reported waveguides lengths $L$ are the optimum terahertz generation lengths. We also report the initial pump intensity $I_{DFG} \equiv P_{u0}/A_{DFG}$ available to the nonlinear process and the initial linear optical light intensity $I_u$ in the pump mode, that is the one to be compared with the optical damage threshold.

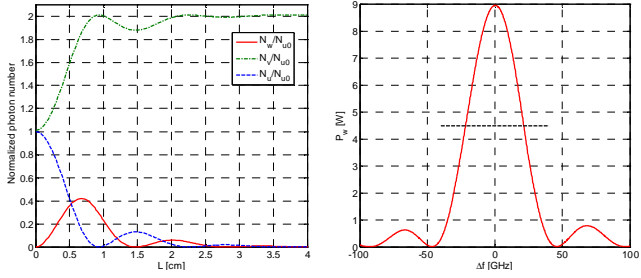

**Fig. 6. Example of photon conversion dynamics and typical generation bandwidth. a)** Simulation of terahertz generation in a GaP waveguide with 100×67.4 μm² cross section, meant to efficiently generate 2 THz radiation. Initial optical pump power and optical signal power are both set to 2 kW. Terahertz absorption damps the typical oscillating behaviour of a lossless parametric amplifier, so that maximum terahertz generation occurs before complete pump depletion. **b)** Terahertz peak power vs. detuning Δ$f$ of the optical signal from the phase matching condition for the InP waveguide of Table 2. The bandwidth is nearly transform limited, as it should be expected in a phase matched process.

Numerical solution of Eq. (1) [47] show that, with a proper length and a nearly optimal waveguide cross-section, the depleted pump regime can be reached launching, for example, optical pump pulses and optical signal pulses with initial peak powers $P_{u0} = P_{v0} = 2$kW and pulse durations in the range 0.1÷10 ns. This can be clearly seen in Fig. 6, showing the photon conversion dynamics vs. propagation length in a GaP waveguide with 100×67.4 μm² cross section. In Table 2 numerical results for different waveguides are presented at the relative optimum conversion lengths, assuming the same initial optical power conditions. By looking at the photon conversion efficiency $\eta_Q$, defined as the ratio between the input pump photon number $N_{u0}$ and the generated terahertz photon number $N_w$, it is clear that, even in case of high terahertz losses, a convenient choice of the waveguide geometry enables to approach the quantum efficiency limit. This is accomplished with pump light intensities well below the damage threshold and pump peak powers orders of magnitude smaller than those used in the bulk experiments reported to date. We stress that the results in Table 2 do not take into account the detrimental effects of nonlinear absorption, cubic nonlinearities and $\chi^{(2)}$ cascading in the optical domain [48], [49], which, in general, are not negligible in centimeters long waveguides with pump intensities in the order of TW/m². Nevertheless, these results can give a flavor of what can be done with modal phase matched zinc-blende crystals, also because the problem of nonlinear absorption can be always overcome by moving to suitably longer optical pump wavelengths. In Fig. 6.b we have also simulated the effect of detuning the optical signal from the phase matching condition, for the InP waveguide of Table 2. As it should be expected for temporally coherent radiation, the predicted bandwidth is nearly transform limited, and it scales inversely with the waveguide length. This means that the optical laser linewidths required to resolve this spectral features must be in the order of a few tens of picometers.

## Conclusion

Using a rigorous numerical mode solver, we have analyzed a set of waveguides made of the most promising nonlinear zinc-blende crystals for phase matched DFG with laser wavelengths around 1.55 μm. Next, from the form invariance of the DFG equations we have been able to synthesize all the physical degrees of freedom involved in the process in just one meaningful number, that is a novel FOM definition that applies to any kind of phase matched DFG process. In this way, we combined linear and nonlinear material properties together with waveguide properties to find the most promising waveguides in the terahertz range and, through numerical simulations, we predicted quantum efficiencies exceeding 40% using pulsed infrared lasers with suitably narrow linewidth and peak power in the order of kWs.

GaP and GaAs are found out to be good materials for modal phase matching, and InP seems to be a very promising material for terahertz applications, if the extrapolated absorption data will be confirmed by experiments. Anyway, our results open novel interesting perspectives for experimental research, not only to provide the missing experimental data and to remove the uncertainties in the nonlinear coefficients data, but also to improve linear and nonlinear material properties of the zincblendes under study, including terahertz absorption, nonlinear optical absorption, optical damage threshold, and second order susceptibility.

Eventually we notice that the presented modal phase matching analysis could be extended to any quadratic nonlinear material, like, for example, the anisotropic crystal GaSe. But, in order to exploit the intrinsic high precision of the proposed method, full characterization of the optical properties is mandatory, including also the determination of the Faust-Henry coefficients for all of the independent elements of the quadratic susceptibility tensor.

## Acknowledgment

M. Cherchi was supported by the Consorzio Nazionale Interuniversitario per le Scienze Fisiche della Materia (CNISM).